\documentstyle[12pt]{article}
\def\beq{\begin{equation}}
\def\eeq{\end{equation}}
\def\barr{\begin{eqnarray}}
\def\earr{\end{eqnarray}}
\def\dag{\dagger}
\def\b{\bigskip}

\def\l{\left(}
\def\r{\right)}
\def\sc{Schr\"odinger }

\begin{document}

\title{Symmetry Breaking in the Schr\"odinger Representation for Chern-Simons
Theories\footnote{UCONN-94-2;hep-th/9406086 \hfill May 1994}}
\author{Gerald Dunne \\
Department of Physics\\
University of Connecticut\\2152 Hillside Road\\
Storrs, CT 06269 USA\\}

\date{}

\maketitle
\begin{abstract}This paper discusses the phenomenon of spontaneous symmetry
breaking in the Schr\"odinger representation formulation of quantum field
theory. The analysis is presented for three-dimensional space-time abelian
gauge theories with either Maxwell, Maxwell-Chern-Simons, or pure Chern-Simons
terms as the gauge field contribution to the action, each of which leads to a
different form of mass generation for the gauge fields.

\end{abstract}
\vskip 0.5 in
\section{Introduction}
\label{sec-intro}
Gauge theories in three-dimensional space-time have the interesting feature
that they permit two different mechanisms for generating masses for gauge
fields. As in four dimensions, the Higgs mechanism, in which a minimally
coupled scalar field acquires a nonzero vacuum expectation value, yields
massive gauge field excitations \cite{abers}. Independently, one may supplement
the usual Maxwell (or Yang-Mills) gauge field action with a Chern-Simon term,
producing a single massive propagating transverse gauge field mode, with mass
determined by the coupling coefficient of the Chern-Simons term \cite{deser1}.
As first considered by Pisarski and Rao \cite{pisarski}, these two
mass-generation techniques may be combined in a Maxwell-Chern-Simons-Higgs
theory, giving two modes with different masses. Finally, one can dispense with
the Maxwell term altogether and consider the Higgs mechanism for gauge theories
with pure Chern-Simons `dynamics', as first discussed by Deser and Yang
\cite{deser2}.

Symmetry breaking is conventionally discussed in a Lagrangian formalism and the
physical degrees of freedom may be identified by a transformation to the
unitary gauge \cite{abers,weinberg}. Perturbative computations, and discussions
of renormalizability, are performed using a gauge fixing prescription, and the
massive gauge field excitations are identified with poles in the gauge field
propagator. The BRS approach to the quantization of gauge theories provides the
most elegant framework for a systematic investigation of more fundamental
canonical issues \cite{kugo}. In this paper, spontaneous symmetry breaking is
analyzed using the Schr\"odinger representation for quantum field theory
\cite{symanzik,jackiw1,keifer}. In this approach, one is concerned with the
Hamiltonian rather than the Lagrangian, and the physical excitations are
identified by a resolution of the constraints in the Weyl gauge, and by a
diagonalization of the quadratic part of the Hamiltonian. For the standard
Higgs mechanism this is a relatively straightforward pedagogical exercise in
the use of the Schr\"odinger representation. However, the introduction of a
Chern-Simons term modifies both the canonical structure and the constraints
\cite{dunne1,asorey}. The quadratic Hamiltonian may still be diagonalized in
the physical subspace, but the explicit diagonalization is remarkably
complicated due to the presence of off-diagonal mixing.

As an immediate corollary of this Schr\"odinger representation analysis we find
a simple explanation of the fact that the two different masses identified by
Pisarski and Rao \cite{pisarski,paul} in the gauge field propagator of
Maxwell-Chern-Simons-Higgs theory coincide precisely with the two different
frequencies arising in the planar quantum mechanical system of charged
particles in a perpendicular magnetic field and a harmonic potential well. This
correspondence is a manifestation of the usual identification of Schr\"odinger
representation quantum field theory with Schr\"odinger representation quantum
mechanics, as has been studied previously in the context of Chern-Simons
theories without symmetry breaking \cite{dunne2}.

This paper is organized as follows. Section 2 contains the Schr\"odinger
representation analysis of Maxwell-Chern-Simons-Higgs theory. This discussion
includes the conventional Higgs mechanism as a special case in which the
Chern-Simons term is removed simply by setting its coupling coefficient to
zero. By diagonalizing the quadratic Hamiltonian in the physical subspace, we
find the two massive scalar modes, with masses in agreement with the
Pisarski-Rao result and in agreement with the quantum mechancal analogy. In
Section 3 the Chern-Simons-Higgs theory is treated in detail, leading to a
single massive mode. Section 4 contains some discussion and an Appendix
contains a constructive method for diagonalizing, with a {\it symplectic}
canonical transformation, a general quadratic Hamiltonian.

\section{Maxwell - Chern - Simons - Higgs Mechanism in the Schr\"odinger
Representation}
\label{sec-mcsh}
In this section we use the Schr\"odinger representation to describe the mass
generation for abelian gauge fields due to the combined effects of the presence
of a Chern-Simons term and the Higgs mechanism. Consider the Lagrange density
\barr
{\cal L}=-{1\over 4e^2}F_{\mu\nu}F^{\mu\nu}-{\kappa\over
2e^2}\epsilon^{\mu\nu\rho}A_\mu\partial_\nu A_\rho -\l D_\mu\phi\r
^{*}D^{\mu}\phi-V
\label{lag}
\earr
where
\beq
\phi\equiv{1\over \sqrt{2}}\l\phi_1 + i \phi_2\r
\eeq
is a complex scalar field, the covariant derivative is
$D_\mu=\partial_\mu+iA_\mu$, $A_\mu$ is an Abelian gauge field, and
$F_{\mu\nu}$ is the corresponding field strength. The space-time metric is
chosen to be ${\rm diag}(-1,1,1)$. The totally antisymmetric $\epsilon$ symbol
is chosen with $\epsilon^{012}=+1$. The scalar field potential $V$ is chosen to
have the standard symmetry-breaking form
\beq
V=\alpha^2{e^2\over 2} \l |\phi |^2 -a^2\r^2
\label{pot}
\eeq
where $a^2$ is some arbitrary mass scale, and $\alpha$ is a dimensionless
constant which controls the ratio of the scalar mass to the gauge mass in the
broken vacuum. Note that $e^2$ has dimensions of mass in $2+1$-dimensional
space-time.

The corresponding momenta for the fields are given by
\barr
\Pi &\equiv& {\partial {\cal L} \over \partial \dot{\phi}} = \l D_0\phi\r
^{*}\cr\cr
\Pi^{*} &\equiv& {\partial {\cal L} \over \partial {\dot{\phi}}^{*}} =
D_0\phi\cr\cr
\pi_i&\equiv&{\partial {\cal L} \over \partial {\dot{A_i}}} = {1\over
e^2}F_{0i}-{\kappa\over 2e^2} \epsilon^{ij}A_j
\label{momenta}
\earr
By a Legendre transformation we find that the Hamiltonian density is
\barr
{\cal H}&\equiv&\Pi \dot{\phi} +\Pi^{*} \dot{\phi}^{*} +\pi_i \dot{A}_i -{\cal
L}\cr\cr
&=&{1\over 2}\Pi_1^2+{1\over 2}\Pi_2^2+{e^2\over 2}\l\pi_i+{\kappa\over
2e^2}\epsilon^{ij}A_j\r ^2 +\l D_i\phi\r^{*} D_i \phi +{1\over
4e^2}F_{ij}F^{ij} +V\cr\cr\cr
&&\quad\quad +A_0\l -\partial_i \pi_i +{\kappa\over
2e^2}\epsilon^{ij}\partial_i A_j - \l\phi_1\Pi_2-\phi_2\Pi_1\r\r
\label{ham}
\earr
The coefficent of $A_0$ in this expression for ${\cal H}$ is the Gauss law
generator which generates fixed-time gauge transformations.

In the spontaneously broken symmetry vacuum the (real) scalar field $\phi_1$
acquires a vacuum expectation value
\beq
\phi_1 \to \phi_1 +\sqrt{2} a
\label{shift}
\eeq
To identify the massive degrees of freedom it is sufficient to consider the
quadratic part of the Hamiltonian density:
\barr
{\cal H}_{\rm quad}&=&{1\over 2}\Pi_1^2+{1\over 2}\Pi_2^2+{e^2\over
2}\l\pi_i+{\kappa\over 2e^2}\epsilon^{ij}A_j\r ^2 +{1\over 2}\l
\partial_i\phi_1\r\l \partial_i \phi_1\r +{1\over 2}\l \partial_i\phi_2\r\l
\partial_i \phi_2\r \cr\cr
&&\quad\quad+{1\over 4e^2}F_{ij}F^{ij}+e^2a^2\phi_1^2+a^2 A_iA_i +\sqrt{2} a
A_i\partial_i\phi_2\cr\cr
&&\quad\quad+A_0\l -\partial_i \pi_i +{\kappa\over 2e^2}\epsilon^{ij}\partial_i
A_j - \sqrt{2} a\Pi_2\r
\label{quadratic}
\earr
{}From this quadratic part of the Hamiltonian it is clear that the $\phi_1$
field (the ``Higgs'' field) separates as a real scalar field of mass
$m_1=\alpha\sqrt{2}ea$. We shall henceforth drop the $\phi_1$ field from our
discussion as it is not relevant to the identification of the remaining massive
modes.

It is now convenient to rewrite the Abelian gauge field $A_i$ using a Hodge
decomposition into its longitudinal and transverse parts. That is, we write
\beq
A_i=\partial_i\lambda -\epsilon^{ij}{\partial_j \over \nabla^2} B
\label{hodge}
\eeq
where $\nabla^2$ is the Laplacian and $B$ is the `magnetic field'
$B=\epsilon^{ij}\partial_iA_j$. Then the gauge field momenta, $\pi_i$, become
\beq
\pi_i=-{\partial_i\over \nabla^2} \Pi_\lambda +\epsilon^{ij}\partial_j\Pi_B
\label{hodgemomenta}
\eeq
In terms of the new fields, $\phi_2$, $\lambda$ and $B$, the quadratic
Hamiltonian density becomes
\barr
{\cal H}_{\rm quad}&=&{1\over 2}\Pi_2^2-{e^2\over 2} \Pi_\lambda {1\over
\nabla^2} \Pi_\lambda -{e^2\over 2}\Pi_B \nabla^2 \Pi_B-{\kappa\over 2}
\nabla^2\lambda \Pi_B +{\kappa\over 2} {1\over \nabla^2} B \Pi_\lambda\cr\cr
&&-{1\over 2}\l\phi_2+\sqrt{2} a\lambda\r \nabla^2 \l\phi_2+\sqrt{2} a\lambda
\r -{\kappa^2\over 8e^2}\lambda\nabla^2\lambda +B\l {1\over 2e^2}-{a^2\over
\nabla^2} - {\kappa^2\over 8e^2\nabla^2}\r B\cr\cr
&&+A_0 \l \Pi_\lambda - \sqrt{2}a \Pi_2+{\kappa\over 2e^2}B\r
\earr
Notice that the fields $\phi_2$ (the ``Goldstone field'') and $\lambda$ (the
longitudinal part of the gauge field) are naturally related in this
Hamiltonian. It is clearly convenient to define new fields
\barr
\chi&=& \phi_2+\sqrt{2} a \lambda\cr
\rho&=& \phi_2-\sqrt{2} a \lambda
\label{new}
\earr
The main advantage of such a transformation is that the Gauss law constraint
now takes the very simple form
\beq
\Pi_\rho = {\kappa\over 4\sqrt{2}ae^2} B
\label{newgauss}
\eeq
In terms of the fields $B$, $\chi$ and $\rho$, the Hamiltonian density becomes
\barr
{\cal H}_{\rm quad}&=&{1\over 2}\Pi_\chi \l 1- {2e^2a^2\over \nabla^2}\r
\Pi_\chi
+{1\over 2}\Pi_\rho \l 1- {2e^2a^2\over \nabla^2}\r \Pi_\rho -{e^2\over 2}
\Pi_B \nabla^2 \Pi_B\cr\cr
&&+\Pi_\chi \l 1+{2e^2a^2\over \nabla^2}\r \Pi_\rho +{\kappa\over 4\sqrt{2}a}
\nabla^2\rho \Pi_B -{\kappa\over 4\sqrt{2}a} \nabla^2\chi \Pi_B +{a\kappa\over
\sqrt{2}\nabla^2}B \Pi_\chi -{a\kappa\over \sqrt{2}\nabla^2}B \Pi_\rho\cr\cr
&&-{1\over 2}\l 1+ {\kappa^2\over 32e^2 a^2}\r \chi \nabla^2 \chi -
{\kappa^2\over 64 e^2a^2}\rho \nabla^2 \rho+B\l {1\over 2e^2}-{a^2\over
\nabla^2} - {\kappa^2\over 8e^2\nabla^2}\r B +{\kappa^2\over 32e^2a^2}\chi
\nabla^2 \rho\cr\cr
&&+A_0 \l -2\sqrt{2} a \Pi_\rho +{\kappa\over 2e^2} B\r
\label{mess}
\earr
This form of the Hamiltonian density looks very complicated. However, the
Hamiltonian simplifies considerably when we now solve the Gauss law constraint
(\ref{newgauss}), as can be done explicitly by expressing a {\it physical}
wavefunctional $\Psi[B,\chi,\rho]$ as
\beq
\Psi[B,\chi,\rho]= exp\l{{i\kappa\over 4\sqrt{2}a e^2}\int B \rho}\r
\psi[B,\chi]
\label{physical}
\eeq
where the wavefunctional $\psi$ is independent of the field $\rho$. This
implies that $\Psi$ automatically satisfies the Gauss law constraint, and we
can derive a new effective Hamiltonian density for the wavefunctional $\psi$ by
acting on $\Psi$ with ${\cal H}_{\rm quad}$.
\barr
{\cal H}_{\rm quad}&=&{1\over 2}\Pi_\chi \l 1- {2e^2a^2\over \nabla^2}\r
\Pi_\chi-{e^2\over 2} \Pi_B \nabla^2 \Pi_B+{a\kappa\over 2\sqrt{2}}\l {3\over
\nabla^2}+{1\over 2a^2e^2}\r B \Pi_\chi -{\kappa\over 4\sqrt{2} a} \nabla^2\chi
\Pi_B\cr\cr
&&+B\l {1\over 2e^2} +{\kappa^2\over 64a^2e^4}-{9\kappa^2\over 32e^2\nabla^2}
-{a^2\over \nabla^2}\r B -{1\over 2}\l 1+{\kappa^2\over 32a^2e^2}\r \chi
\nabla^2 \chi
\label{offdiag}
\earr
Notice that all dependence on the field $\rho$ has disappeared from the
Hamiltonian density. This serves as a nontrivial check that the Gauss law has
been solved correctly. This is also the Schr\"odinger representation version of
the usual statement that the longitudinal component $\lambda$ of the gauge
field ``eats'' the Goldstone boson $\phi_2$ - they combine into the fields
$\rho$ and $\chi$ as in (\ref{new}) and one of these fields, $\rho$, disappears
when Gauss' law is solved.

The final result is that we are left with a quadratic Hamiltonian density
involving just the fields $B$ and $\chi$, which we can therefore view as (in a
sense to be made precise below) the ``physical'' fields.

To truly identify the two independent physical fields, we first normalize the
fields so that the kinetic parts of the Hamiltonian density take the form
${1\over 2}({\rm momentum})^2$. Define the rescaled fields
\barr
B&\equiv&\sqrt{-e^2\nabla^2} \;\tilde{B}\cr
\chi&\equiv&\sqrt{1-{2e^2a^2\over \nabla^2}}\;\tilde{\chi}
\label{rescale}
\earr
In terms of these rescaled fields, the Hamiltonian density is
\barr
{\cal H}_{\rm quad}&=&{1\over 2} \Pi_{\tilde{\chi}}^2+{1\over 2}
\Pi_{\tilde{B}}^2
+{\kappa\over 4\sqrt{2}ae}\sqrt{2e^2a^2-\nabla^2}\tilde{\chi} \Pi_{\tilde{B}}
-{\kappa\over 4\sqrt{2}ae} {{6a^2e^2+\nabla^2}\over
\sqrt{2a^2e^2-\nabla^2}}\tilde{B} \Pi_{\tilde{\chi}}\cr\cr
&&+{1\over 2}\l 1+{\kappa^2\over 32a^2e^2}\r \tilde{\chi} \l 2 a^2e^2-\nabla^2
\r \tilde{\chi}\cr\cr
&&+{1\over 2} \tilde{B}\l 2a^2e^2 \l 1 +{9\kappa^2 \over 32a^2e^2}\r - \l
1+{\kappa^2\over 32a^2e^2}\r\nabla^2 \r \tilde{B}
\label{newham}
\earr
Notice that in the absence of a Chern-Simons term (i.e. set $\kappa=0$), this
Hamiltonian simplifies dramatically to
\barr
{\cal H}_{\rm quad}|^{\kappa=0}={1\over 2} \Pi_{\tilde{\chi}}^2+{1\over 2}
\Pi_{\tilde{B}}^2 +{1\over 2}\tilde{\chi} \l 2 a^2e^2-\nabla^2 \r \tilde{\chi}
+{1\over 2} \tilde{B}\l 2a^2e^2  - \nabla^2 \r \tilde{B}
\label{free}
\earr
from which we clearly see that the fields $\tilde{B}$ and
$\tilde{\chi}$ represent scalar degrees of freedom with equal masses $\sqrt{2}
ae$, to be compared with the Higgs field ($\phi_1$) mass, $\alpha \sqrt{2}ae$.
The special choice of $\alpha=1$ in the potential (\ref{pot}) makes the gauge
and Higgs masses degenerate and corresponds to the self-dual point of the
Abelian Higgs model \cite{bog}.

When $\kappa\neq 0$, there are cross-terms in the Hamiltonian (\ref{newham})
which mix the fields and momenta, thereby complicating the direct
identification of the masses of the elementary fields. The general technique
for diagonalizing such a Hamiltonian density is described in detail in the
Appendix. One can consider the quadratic quantum mechanical Hamiltonian of the
form
\barr
H&=&{1\over 2}p_1^2+{1\over 2}p_2^2+b_1 q_1p_2+b_2q_2p_1+{1\over 2}c_1^2
q_1^2+{1\over 2}c_2^2 q_2^2\cr\cr
&\equiv&{1\over 2}\xi^T h \xi
\label{qmech}
\earr
where in the second line we have written $\xi$ as a phase-space vector
$\xi=(p_1, p_2, q_1, q_2 )$, and $h$ is the $4\times 4$ real symmetric matrix
\beq
h = \l \begin{array}{cccc}
1&0&0&b_2\\0&1&b_1&0\\0&b_1&c_1^2&0\\b_2&0&0&c_2^2\end{array}\r
\eeq
In order to identify the normal mode frequencies of such a Hamiltonian we need
to be able to diagonalize $h$ with a {\it symplectic} (rather than an {\it
orthogonal}) matrix.

 Since the Hamiltonian equations of motion for a quadratic Hamiltonian such as
(\ref{qmech}) are {\it linear} and have the form
\beq
\dot{\xi} = -{\cal E} h \xi
\label{eqs}
\eeq
where the $4\times 4$ matrix ${\cal E}$ is
\beq
{\cal E} = \l \begin{array}{cccc}
0&0&1&0\\0&0&0&1\\-1&0&0&0\\0&-1&0&0\end{array}\r
\eeq
the eigenmodes of the quadratic Hamiltonian (\ref{qmech})
are given by the eigenvalues of the matrix $i {\cal E} h$. A simple
calculation gives the squares of the eigenvalues of $i {\cal E} h$ as
\beq
\omega^2_\pm = {1\over 2} \l c_1^2+c_2^2-2 b_1b_2 \r \pm {1\over 2}\sqrt{\l
c_1^2+c_2^2-2 b_1b_2 \r ^2 -4\l c_1^2-b_1^2\r \l c_2^2 -b_2^2 \r}
\label{modes}
\eeq
We can now apply this result directly to the hamiltonian density (\ref{newham})
with phase-space fields $\xi=\l \Pi_{\tilde{\chi}}, \Pi_{\tilde{B}},
\tilde{\chi}, \tilde{B}\r$. Then the normal modes of the hamiltonian
(\ref{newham}) are given by Equation (\ref{modes}) as
\beq
\omega^2_\pm=-\nabla^2+2a^2e^2+{\kappa^2\over 2} \pm {\kappa\over 2}
\sqrt{\kappa^2+8a^2 e^2}
\label{answer}
\eeq
It is worth commenting on the remarkable simplicity of this result, especially
considering the very complicated form of the Hamiltonian (\ref{newham}). Notice
that each $\omega_\pm^2$ is positive, and that each has the form
$-\nabla^2+(mass)^2$. This clearly identifies the Hamiltonian as one describing
two scalar fields, with masses (squared) given by
\beq
m^2_\pm=2a^2e^2+{\kappa^2\over 2} \pm {\kappa\over 2} \sqrt{\kappa^2+8a^2 e^2}
\eeq
which agrees with the masses found by Pisarski and Rao \cite{pisarski} in their
analysis of the covariant gauge propagator. This result can alternatively be
expressed as
\beq
m_\pm = {\kappa \over 2}\l\sqrt{1+8{a^2e^2\over \kappa^2}} \pm 1\r.
\label{masses}
\eeq
The linear canonical (symplectic) transformation used to find these normal
modes of the Hamiltonian density (\ref{newham}) transforms the quadratic
Hamiltonian density into its diagonalized oscillator form:
\beq
{\cal H}_{\rm quad} = {1\over 2}\sqrt{-\nabla^2+m_+^2} \l a_+^{\dag} a_+ +a_+
a_+^{\dag}
\r + {1\over 2}\sqrt{-\nabla^2+m_-^2} \l a_-^{\dag} a_- +a_- a_-^{\dag} \r
\eeq
where $a_\pm$ are annihilation operators for two independent scalar fields. The
explicit linear transformation relating the phase space fields
$\Pi_{\tilde{\chi}}$, $\Pi_{\tilde{B}}$, $\tilde{\chi}$ and  $\tilde{B}$ with
the oscillator fields $a_\pm$ and $a_\pm^{\dag}$ is extremely complicated and
is described in the Appendix. These oscillators satisfy the canonical
commutation relations
\barr
[a_\pm (\vec{x}), a^\dag_\pm(\vec{y})] &=& \delta (\vec{x}-\vec{y}) \cr\cr
[a_\pm (\vec{x}), a^\dag_\mp(\vec{y})]= &0& = [a_\pm (\vec{x}), a_\mp(\vec{y})]
\earr
Re-introducing the Higgs field, $\phi_1$, the ground state wavefunctional in
the symmetry breaking vacuum is
\barr
\Psi_0 [\phi_+, \phi_-, \phi_1] = {\det}^{1/4} \l {\omega_+ \omega_-
\omega_1\over \pi^3}  \r exp \left[ -{1\over 2} \int \l \phi_+ \omega_+ \phi_+
+ \phi_- \omega_- \phi_- +\phi_1 \omega_1 \phi_1\r\right]
\earr
where $\omega_\pm=\sqrt{-\nabla^2+m_\pm^2}$ ,
$\omega_1=\sqrt{-\nabla^2+2\alpha^2e^2a^2}$, and $\phi_\pm$ are \sc
representation fields given by
\barr
\phi_\pm={1\over \sqrt{2 \omega_\pm}}\l a_\pm + a^\dag_\pm \r
\earr

The result (\ref{masses}) for the masses in the broken phase may also be
understood in the context of Chern-Simons quantum mechanics, in which many
features of an Abelian topologically massive gauge theory may be
analyzed in an analogue quantum mechanical model describing the planar motion
of a particle in a uniform perpendicular magnetic field \cite{dunne2}. The
Chern-Simons coefficient ${\kappa\over e^2}$ plays the role of the external
magnetic field strength, and the Chern-Simons modifications to the equations of
motion behave like the Lorentz force modifications to the free-particle
equations of motion. Extending this analogy to include a Higgs-like mass term
$a^2 e^2 A_i A_i$ for the gauge fields is like adding an additional isotropic
harmonic potential to the analogue quantum mechanical model. Indeed, the
two-dimensional quantum mechanical Hamiltonian
\beq
H={1\over 2}\l p_i+{\kappa\over 2e^2}\epsilon^{ij}q_j\r ^2 +a^2 e^2 q_iq_i
\label{mag}
\eeq
is well known to be separable into two independent harmonic oscillator
Hamiltonians of frequencies
\beq
\omega_\pm={\kappa \over 2}\l\sqrt{1+8{a^2e^2\over \kappa^2}} \pm 1\r
\eeq
which agree precisely with the masses in (\ref{masses}). This comparison is due
to the standard reduction of Schr\"odinger representation quantum field theory
to Schr\"odinger representation quantum mechanics in a derivative expansion.
The field theoretical frequencies $\omega_\pm^2 =-\nabla^2 +m_\pm^2$ reduce to
$m_\pm^2$, and the Lagrange density reduces to the corresponding quantum
mechanical Lagrangian.

\b
\section{Chern-Simons-Higgs Mechanism in the Schr\"odinger Representation}
\label{sec-csh}
In this section, we discuss the Schr\"odinger representation implementation of
the Higgs mechanism in the presence of a Chern-Simons, but {\it without} a
Maxwell kinetic term for the gauge field. This ``pure Chern-Simons'' Higgs
mechanism was first discussed by Deser and Yang \cite{deser2}. It warrants
separate investigation in the Schr\"odinger representation because, as is well
known, the canonical structure is very different without a Maxwell kinetic term
for the gauge field.

Consider the Lagrange density
\barr
{\cal L}=-{\kappa\over 2e^2}\epsilon^{\mu\nu\rho}A_\mu\partial_\nu A_\rho -\l
D_\mu\phi\r^{*}D^{\mu}\phi-V
\label{purelag}
\earr
where the scalar field potential is as defined in (\ref{pot}). The scalar field
canonical momenta, $\Pi$ and $\Pi^*$, are as in (\ref{momenta}). However, since
the Lagrange density (\ref{purelag}) is first-order in time derivatives, the
gauge fields $A_1$ and $A_2$ are canonically conjugate to one another
\cite{jackiw2,dunne1}. One is free to choose a convenient polarization, and
here I choose to identify $A_1$ as a coordinate field, with $A_2$ as the
corresponding momentum field:
\barr
\pi_1 = -{\kappa\over e^2} A_2
\earr
The Hamiltonian density now reads
\barr
{\cal H} &=&\Pi \dot{\phi}+\Pi^* \dot{\phi}^* +\pi_1 \dot{A}_1 -{\cal L}\cr
&=&\Pi^* \Pi +(D_i\phi)^* D_i\phi +V +A_0 \l {\kappa\over
e^2}\epsilon^{ij}\partial_i A_j +i\l\phi^*\Pi^*-\phi\Pi\r\r
\earr
where we have dropped a total time derivative term \cite{dunne1} which does not
affect the identification of the physical modes.

In the spontaneously broken vacuum, we shift the scalar field as in
(\ref{shift}), and (just as before) the real Higgs scalar field $\phi_1$
separates from the rest of the quadratic Hamiltonian density, identifying
itself as a real scalar field of mass $\alpha\sqrt{2} a e$. The remainder of
the quadratic Hamiltonian density is
\barr
{\cal H}_{\rm quad} = {1\over 2}\Pi_2^2 +{1\over
2}\l\partial_i\phi_2\r\l\partial_i\phi_2\r+a^2A_iA_i+\sqrt{2}aA_i\partial_i\phi_
2+A_0\l {\kappa\over e^2}\epsilon^{ij}\partial_iA_j-\sqrt{2} a\Pi_2\r
\earr
We now make the Hodge decomposition for the gauge field as in (\ref{hodge}),
but note that this is now to be viewed as a {\it linear} transformation of the
{\it phase space} fields $A_1$ and $A_2$ to the longitudinal and transverse
fields $\lambda$ and $B$, respectively. Thus, $B$ and $\lambda$ are now to be
viewed as conjugate fields, and we can choose $\lambda$ as the coordinate
field, in which case we identify $B$ as the corresponding momentum
\barr
B={e^2\over \kappa} \Pi_{\lambda}
\earr
In terms of these fields, the quadratic Hamiltonian density becomes
\barr
{\cal H}_{\rm quad} = {1\over 2}\Pi_2^2 -{a^2e^4\over \kappa^2}
\Pi_{\lambda}{1\over \nabla^2} \Pi_{\lambda}-{1\over
2}\l\phi_2+\sqrt{2}a\lambda\r\nabla^2\l\phi_2+\sqrt{2}a\lambda\r+A_0\l
\Pi_{\lambda}-\sqrt{2} a\Pi_2\r
\earr
In this form of the Hamiltonian we recognize once again the field combinations
$\chi=\phi_2+\sqrt{2}a\lambda$ and $\rho=\phi_2-\sqrt{2}a\lambda$, defined
before in (\ref{new}). With these fields, the Gauss law constraint takes the
simple form
\barr
\Pi_\rho =0
\earr
Thus, when acting on physical state wavefunctionals
$\Psi[\chi,\rho]=\Phi[\chi]$ which are annihilated by $\Pi_{\rho}$, the
effective quadratic Hamiltonian density is
\barr
{\cal H}_{\rm quad}={1\over 2}\Pi_\chi \l 1-{4a^4e^4\over \kappa^2\nabla^2} \r
\Pi_\chi -{1\over 2} \chi\nabla^2\chi
\earr
To identify the physical mode the field $\chi$ is rescaled as
$\chi=\sqrt{1-{4a^4e^4\over \kappa^2\nabla^2}} \tilde{\chi}\;$, in which case
the quadratic Hamiltonian density becomes
\barr
{\cal H}_{\rm quad} = {1\over 2} \Pi_{\tilde{\chi}}^2 +{1\over 2}
\tilde{\chi}\l -\nabla^2+{4a^4e^4\over \kappa^2}\r \tilde{\chi}
\earr
which clearly identifies the single scalar mode of mass
\barr
m={2a^2e^2\over \kappa},
\earr
in agreement with the result of Deser and Yang \cite{deser2}.

To conclude this section, we note that it is possible to arrive at this mass
from the results of the previous section by considering the limit in which the
Maxwell term is removed from the Lagrangian density (\ref{lag}). Comparing the
Lagrange densities (\ref{lag}) and (\ref{purelag}), we see that formally one
can arrive at the pure Chern-Simons case via the limit
\barr
e^2&\to&\infty\cr
\kappa&\to&\infty\cr
{\rm with}& {\kappa\over e^2}& {\rm fixed}
\earr
In this limit the masses $m_\pm$ in (\ref{masses}) behave as
\barr
m_+ &\to &\infty\cr
m_- &\to & {2a^2e^2\over \kappa}
\earr
so that excitations associated with $m_+$ decouple, leaving the single mode of
mass $m=2a^2e^2/\kappa$. This is precisely analogous to the lowest Landau level
projection in the analogue quantum mechanical problem in which the external
magnetic field strength goes to infinity and the resulting dynamics is governed
by the frequency of the harmonic well \cite{girvin,dunne2}.
\b
\section{Conclusions}
\b
This paper presents a pedagogical exercise in implementing the field theoretic
\sc representation for gauge theories in $2+1$ dimensions involving mass
generation effects via the Higgs mechanism and/or Chern-Simons terms. The
presence of a Chern-Simons term introduces off-diagonal mixing in the quadratic
Hamiltonian, a feature which complicates the diagonalization procedure in an
interesting manner. The \sc representation approach explains why the resulting
two massive modes have masses corresponding to the two characteristic
frequencies of an analogue quantum mechanical model of planar particles in a
perpendicular magnetic field and a harmonic potential well. This also shows
that the relationship between the pure Chern-Simons-Higgs theory, with one
massive gauge mode, and the Maxwell-Chern-Simons-Higgs theory, with two massive
gauge modes, is one of a truncation of the Hilbert space as one mode decouples
to infinity, analogous to the projection of dynamics onto the lowest Landau
level in the quantum mechanical model. Even in the quantum mechanical case,
there are well-known subtleties \cite{girvin,dunne2} involved with the
combination of such a truncation of the Hilbert space with the development of a
perturbation expansion. In field theoretical models, such as
Maxwell-Chern-Simons-Higgs theory or Chern-Simons-Higgs theory, these
subtleties are compounded by regularization concerns. Nevertheless, this \sc
representation analysis clearly identifies this phenomenon as the physical
source of various delicate issues that have been found in the analysis of
Chern-Simons theories with symmetry breaking using more conventional covariant
perturbation theory \cite{khlebnikov,leblanc,khare}, where it has been found
that the computation of loop effects does not necessarily commute with the
taking of such projections.

\vfill
\eject
\section{Appendix}
\b
In this appendix I present the details of the symplectic diagonalization of the
Hamiltonian
\beq
H={1\over 2}\xi^{\rm T} h \xi
\label{h}
\eeq
where $\xi\equiv\l p_1, p_2, q_1, q_2 \r$ and $h$ is the real symmetric
$4\times 4$ matrix
\beq
h = \l \begin{array}{cccc}
1&0&0&b_2\\0&1&b_1&0\\0&b_1&c_1^2&0\\b_2&0&0&c_2^2\end{array}\r.
\label{hh}
\eeq
That is, we shall seek a real matrix ${\cal S}$ such that
\beq
h={\cal S}^{\rm T} \l \begin{array}{cccc}
1&0&0&0\\0&1&0&0\\0&0&\omega_+^2&0\\0&0&0&\omega_-^2\end{array}\r {\cal S}
\label{diagonal}
\eeq
with ${\cal S}$ being {\it symplectic} (not orthogonal!)
\beq
{\cal S} {\cal E} {\cal S}^{\rm T} = {\cal E}
\eeq
and where $\omega_\pm$ are the eigenvalues of the matrix $i{\cal E}h$. The
condition that ${\cal S}$ be symplectic is required so that the canonical
structure of the phase space variables is preserved by the linear
transformation $\xi\to\xi^\prime ={\cal S}\xi$.

In the familiar special case (\ref{mag}) of two-dimensional motion of a
particle in a {\it uniform} magnetic field and an {\it isotropic} harmonic
potential, the matrix $h$ has the form
\beq
h= \l \begin{array}{cccc} 1&0&0&-{B\over 2}\\0&1&{B\over 2}&0\\0&{B\over
2}&\Omega^2&0\\-{B\over 2}&0&0&\Omega^2\end
{array}\r
\label{special}
\eeq
where $B$ is the magnetic field strength, $\Omega^2\equiv\omega^2+{B^2\over
4}$, and $\omega$ is the frequency of the isotropic harmonic potential. Then
the eigenvalues of $i{\cal E}h$ are
\beq
\omega_\pm=\Omega\pm{B\over 2}
\eeq
and the real symplectic matrix ${\cal S}$ which diagonalizes $h$ is
\beq
{\cal S}=\l \begin{array}{cccc} \sqrt{{\omega_+\over
2\Omega}}&0&0&-\sqrt{{\Omega\omega_+\over 2}}\\-\sqrt{{\omega_-\over
2\Omega}}&0&0&-\sqrt{{\Omega\omega_-\over 2}}\\0&{1\over
\sqrt{2\Omega\omega_+}}&\sqrt{{\Omega\over 2\omega_+}}&0\\0&{1\over
\sqrt{2\Omega\omega_-}}&-\sqrt{{\Omega\over 2\omega_-}}&0\end{array}\r
\label{sim}
\eeq
The linear transformation $\xi\to\xi^\prime={\cal S}\xi$ with this form of
${\cal S}$ is the familiar one \cite{dunne2} which separates this model into
two independent
harmonic oscillators of frequency $\omega_\pm=\Omega\pm{B\over 2}$. However, in
the general case, with $h$ given by (\ref{hh}), it is considerably more
complicated to construct the symplectic matrix ${\cal S}$ which diagonalizes
$h$. To introduce the general method for constructing such an ${\cal S}$,
consider the equations of motion arising from the Hamiltonian (\ref{h}):
\beq
\dot{\xi} = - {\cal E} h \xi
\eeq
This shows that the normal modes of this Hamiltonian system are given by the
eigenvalues of the matrix $i {\cal E} h$. Further, as outlined below, the
symplectic matrix ${\cal S}$ is constructed from the eigenvectors of $i {\cal
E} h$.

Note first of all that the eigenvalues of $i {\cal E} h$ are {\it real}, and
occur in $\pm$ pairs. This is because
\barr
\det \l i{\cal E} h - \omega I \r &=&\det \l \l i{\cal E} h - \omega I \r ^{\rm
T} \r \cr
&=&\det \l -ih{\cal E} - \omega I \r \cr
&=&\det \l i{\cal E} h + \omega I \r
\earr
Further, if $v$ is an eigenvector of $i {\cal E} h$ with eigenvalue $\omega$,
then $v^{*}$ is an eigenvector with eigenvalue $-\omega^{*} =-\omega$. Now let
$V$ be the $4\times 4$ matrix whose columns are formed by the eigenvectors of
$i {\cal E} h$. Then these columns may be normalized in such a way that
\beq
V^{\dag} {\cal E} V = i \l \begin{array}{cccc}
-1&0&0&0\\0&-1&0&0\\0&0&1&0\\0&0&0&1\end{array}\r
\label{inner}
\eeq
In particular, note that eigenvectors of $i {\cal E} h$ with {\it different}
eigenvalues are orthogonal with respect to the inner product $w^{\dag}{\cal
E}v$. Using equation (\ref{inner}) we see that the equations of motion for
$\xi$ may be written as
\barr
V^{\dag}{\cal E} \dot{\xi}&=&-i \l V^{\dag} h V \r \l \begin{array}{cccc}
-1&0&0&0\\0&-1&0&0\\0&0&1&0\\0&0&0&1\end{array}\r V^{\dag} {\cal E}\xi\cr\cr
&=&i\l \begin{array}{cccc}
\omega_+&0&0&0\\0&\omega_-&0&0\\0&0&-\omega_+&0\\0&0&0&-\omega_-\end{array}\r
V^{\dag} {\cal E} \xi
\earr
These equations of motion are now in the diagonal ``oscillator'' form
\barr
\dot{a}_\pm ^{\dag} &=& i \omega_\pm a_\pm ^{\dag}\cr
\dot{a}_\pm &=& -i \omega_\pm a_\pm
\earr
which would follow from the Hamiltonian
\beq
H={\omega_+ \over 2} \l a_+^{\dag}a_+ + a_+ a_+^{\dag} \r + {\omega_- \over 2}
\l a_-^{\dag}a_- + a_- a_-^{\dag} \r
\eeq
So we see that the transformation from the initial phase space variables
$\xi\equiv\l p_1, p_2, q_1, q_2 \r$ to the oscillators $a_\pm$ and
$a_\pm^{\dag}$ is
\beq
\l \begin{array}{c}
a_+^{\dag}\\a_-^{\dag}\\a_+\\a_-\end{array}\r = V^{\dag}{\cal E}\l
\begin{array}{c}
p_1\\p_2\\q_1\\q_2\end{array}\r \equiv V^{\dag}{\cal E}\xi
\eeq
We may alternatively wish to present the diagonalized Hamiltonian in terms of
new canonical coordinates and momenta, $q_\pm$ and $p_\pm$, related to the
oscillators $a_\pm$ and $a_\pm^{\dag}$ as
\barr
q_\pm&\equiv&\pm{1\over \sqrt{2\omega_\pm}} \l a_\pm + a_\pm^{\dag} \r\cr
p_\pm&\equiv&\mp i \sqrt{{\omega_\pm \over 2}} \l a_\pm - a_\pm^{\dag} \r
\earr
These coordinates and momenta are thus related to the original ones by the net
transformation
\beq
\l \begin{array}{c}
p_+\\p_-\\q_+\\q_-\end{array}\r = {\cal S}\l \begin{array}{c}
p_1\\p_2\\q_1\\q_2\end{array}\r
\eeq
where ${\cal S}$ is the matrix
\beq
{\cal S}=\l \begin{array}{cccc} i\sqrt{{\omega_+\over
2}}&0&-i\sqrt{{\omega_+\over 2}}&0\\0&-i\sqrt{{\omega_-\over
2}}&0&i\sqrt{{\omega_-\over 2}}\\{1\over \sqrt{2\omega_+}}&0&{1\over
\sqrt{2\omega_+}}&0\\0&-{1\over \sqrt{2\omega_-}}&0&-{1\over
\sqrt{2\omega_-}}\end{array}\r V^{\dag} {\cal E}
\label{symplectic}
\eeq
By construction, this matrix ${\cal S}$ is a real symplectic matrix.
Furthermore, by construction, it diagonalizes $h$ in the sense of
(\ref{diagonal}).

Note that this result is completely general, for any Hamiltonian of the form
(\ref{qmech}). However, for such a general Hamiltonian the explicit expression
for the normalized eigenvectors of $i{\cal E}h$ (recall that these eigenvectors
form the columns of the matrix $V$) is very complicated - note that even the
expression for the {\it eigenvalues} (\ref{modes}) is quite involved. It is,
however, instructive to illustrate this procedure in the special case
(\ref{special}). Here the matrix $V$ of eigenvectors of $i{\cal E} h$ is
\beq
V={\sqrt{\Omega}\over 2}\l \begin{array}{cccc}
1&1&1&1\\{-i}&{i}&{i}&-{i}\\-{i\over \Omega}&-{i\over \Omega}&{i\over
\Omega}&{i\over \Omega}\\-{1\over \Omega}&{1\over \Omega}&-{1\over
\Omega}&{1\over \Omega}\end{array}\r
\eeq
It is straightforward to check that this matrix $V$ is correctly normalized, as
in (\ref{inner}), and that using it in equation (\ref{symplectic}) yields the
correct real symplectic diagonalizing matrix (\ref{sim}).


\end{document}